# A Proposed Improvement Equalizer for Telephone and Mobile Circuit Channels

Dr. Laith Awda Kadhim
Al-Khwarizmi College Of Engineering, Mechatronics Engineering Department
University Of Baghdad
Baghdad – Iraq
Postdoctoral in (ICT), CRP uOttawa
Ottawa–Canada
laithawda@yahoo.com

Dr. Salih Mohammed, Osamah Saad
Electrical Department
University of Anbar
Baghdad – Iraq
dr_salih_moh@yahoo.com

*Abstract*— **In the transmission of digital data at a relatively high rate over a particular band limited channel, it is normally necessary to employ an equalizer at the receiver in order to correct the signal distortion introduced by the channel .ISI (inter symbol interference) leads to large error probability if it is not suppressed .The possible solutions for coping with ISI such as equalization technique. Maximum Likelihood Sequence Estimation (MLSE) implemented with Viterbi algorithm is the optimal equalizer for this ISI problem sense it minimizes the sequence of error rate. This estimator involves a very considerable amount of equipment complexity especially when detecting a multilevel digital signal having large alphabet, and/or operating under a channel with long impulse response, this arises a need to develop detection algorithms with reduced complexity without losing the performance. The aim of this work is to study the various ways to remove the ISI, concentrating on the decision-based algorithms (DFE, MLSE, and near MLSE), analyzing the difference between them from both performance and complexity point of view. An Improved non linear equalizer with Perturbation algorithm has been suggested which trying to enhance the performance and reduce the computational complexity by comparing it with the other existing detection algorithms.**

## I. INTRODUCTION

For high-rate digital transmission systems, a major difficulty with signaling is the increased amount of intersymbol interference (ISI) [1] [2]. Performance of the symbol-by symbol detector (SSD) becomes unsatisfactory since this form of interference cannot be tackled by simply raising the signal power. Early approach uses equalization technique [3]. Equalization techniques fall into two broad categories: linear and nonlinear. However, linear equalization techniques typically suffer from noise enhancement on frequency-selective fading channels, and are therefore not used in most wireless applications. Among nonlinear equalization techniques, decision-feedback equalization (DFE) is the most common, since it is fairly simple to implement and does not suffer from noise enhancement. The optimal equalization technique to use is MLSE. The Viterbi algorithm can be used for MLSE; the complexity of this equalization technique grows exponentially with the channel delay spread [4]. Near Maximum Likelihood Detection method can be used for channels with long sampled impulse response and a modulation technique with many symbol levels, where the Viterbi-algorithm detector becomes impractical [5].

## II. PROPOSED MODEL FOR THE NON LINEAR EQUALIZER

In this paper, a new approach is presented for 16-QAM channel equalization where the detection on the data symbol is delayed for one and two sample. The method is suboptimal because it selects a low number of states than MLSE but can achieve good bit-error ratio (BER) performance with low computational complexity. The algorithm iteratively minimizes the Euclidean distance between the detected and received signal sequences. The detected hard symbols are then used to cancel ISI when computing the n-delay form of the symbol (bit) probabilities for soft decision decoding. This paper compares the simulation results of improved nonlinear equalizer by using perturbation algorithm with the other reduced complexity algorithms under the same conditions.

## III. IMPROVED NONLINEAR EQUALIZER

The conventional nonlinear equalizer is the arrangement of Fig.1 in which the delay in the detection is zero, at the output of the adaptive linear filter the detected data symbol $s'_k$ is determined from the received signal $r_k$.

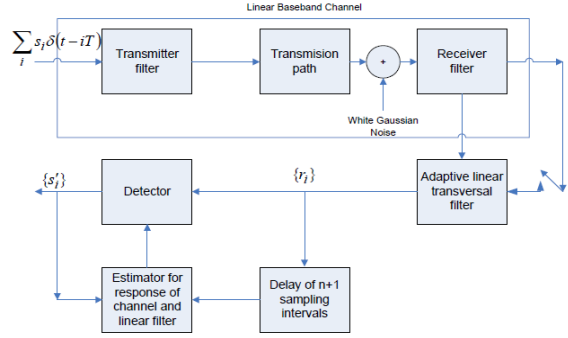

Fig. 1. Data transmission system

The estimator for the response of the channel and adaptive filter can be shown by [6]:

$$f'_k = \sum_{h=1}^{g} s'_{k-h} y_h \quad (1)$$

This estimator forms the estimate of the intersymbol interference in $r_k$, to give the equalized signal:

$$e_k = r_k - f'_k \quad (2)$$

This is fed to a simple threshold –level detector. With the correct detection of the data symbol:

$$s_{k-1}, s_{k-2}, \ldots, s_{k-g} \quad (3)$$

The equalized signal becomes:

$$e_k = s_k + w_k \quad (4)$$

Since $y_0 = 1$, so that $s_k$ is detected without intersymbol interference. $s'_k$ is now taken as its possible value closest to $e_k$.

## IV. IMPROVED NONLINEAR EQUALIZER WITH PERTURBATION OF ONE SAMPLE DELAY

The algorithm works on a first case where the detection of the data symbol $s_k$ is delayed until the receipt of $r_{k+1}$. the detector now has determined $s'_{k+1}$, $s'_{k+2}$, ....., so that it can form the two equalized signals $e_k$ [7].

$$e_{k,k+1} = r_{k+1} - \sum_{h=1}^{g-1} s'_{k-h} y_{h+1} \quad (5)$$

We now propose a two-step optimization procedure to find the symbol vector that will minimize the sequence error $e^2$.

Step one: Let $s' = 0$. Starting $s'_k$ with k=20, select 16-QAM alphabet hard symbol $s'_k$ that yields the least value for $\left\| r(k) - y_0 s'_k - \sum_{h=1}^{g} y_h s'_{k-h} \right\|^2$.

Repeat the search for the next symbol $s'_{k+1}$ using previously detected symbol $s'_k$,. The result of this hard symbol search is $s'^*$.

Step Two: Using the obtained $s'^*$ as an initial solution, starting from k=0 test the nearest neighbors of $s'^*_k$ (neighbors of $s'^*_k$ are 2 or 3 or 4 of 16-QAM constellation) in the 16-QAM constellation for a possible reduction in the sequence error $e^2$. If the sequence error is reduced (or reduced most) by one of the neighbors, then update $s'^*_k$ with that neighbor. Repeat the procedure for the next symbol $s'^*_{k-2}$ and so on until k is reached to the last value. This updated $s'^*$ is then the final estimate of the of the transmitted hard-symbol vector in the sequence sense denoted by $s'_k$ sequence [8], with the correct values of $s'_{k-1}$, $s'_{k-2}$ ,...., $s'_{k-g}$, $e_k$ is given by

$$e_{k,k+1} = s_{k+1} + s_k y_1 + w_{k+1} \qquad (6)$$

where $y_0 = 1$.

In this system, $s'_k$ is taken as its possible value $x_k$ for which the cost

$$c_{1,k} = |e_k - x_k|^2 + |e_{k+1} - x_k y_1 - x_{k+1}|^2 \qquad (7)$$

is minimum over all combination of possible value of $x_k$ and $x_{k+1}$. This involve the computation of only 16 different values of $c_{1,k}$ since ,for any one value of $x_k$, the real and imaginary parts of the value of $x_{k+1}$ that minimizes $c_{1,k}$ are determined by a process of simple threshold comparison. Operating separately on the real and imaginary parts of $e_{k,k+1} - x_k y_1$. In this system, the detector determines first, by simple threshold comparison, the value of $x_k$ that has least cost, and for this value of $x_k$ it selects the value of $x_{k+1}$ that has least cost. For these selected values of $x_k$, and $x_{k+1}$, are given by all combinations of the real and imaginary parts, and more than not correspond to the smallest values $|e_k - x_k|^2$. Then test the neighbors $x_k$, and $x_{k+1}$. Now for each of these values of $x_k$, the value of $x_{k+1}$ that minimize $c_{1,k}$ is determined by simple threshold comparison. The detector now evaluates $c_{1,k}$ for each of the 16 combinations of values of $x_k$ and $x_{k+1}$, and then takes as the detected value $s'_k$ of the data symbol $s_k$ the value of $x_k$ corresponding to the smallest $c_{1,k}$.

V. IMPROVED NONLINEAR EQUALIZER WITH PERTURBATION OF TWO SAMPLE DELAY

The next step in this algorithm, the detection of the data symbol $s_k$ is delayed until the receipt of $r_{k+2}$. The detector now forms the three equalized signals $e_k$, $e_{k,k+1}$ and $e_{k,k+2}$, where $e_k$ and $e_{k,k+1}$ are given by

$$e_{k,k+2} = r_{k+2} - \sum_{h=1}^{g-2} s'_{k-h} y_{h+2} \qquad (8)$$

As mentioned above with the detection of the data symbol $s_k$ is delayed until the receipt of $r_{k+1}$, the same steps for the data symbol $s_k$ is delayed until the receipt of $r_{k+2}$ are repeated to find the symbol vector that will minimize the sequence error $e^2$.

This updated $s'^*$ is then the final estimate of the of the transmitted hard-symbol vector in the sequence sense denoted by $s'_k$ sequence. With the correct values of $s'_{k-1}$, $s'_{k-2}$....$s'_{k-g}$ the equalized signals $e_k$ and $e_{k,k+1}$

$$e_{k,k+2} = s_{k+2} + s_{k+1} y_1 + s_k y_2 + w_{k+2} \qquad (9)$$

$s'_k$ is taken as its possible value $x_k$ for which the cost

$$c_{2,k} = c_{1,k} + |e_{k,k+2} - x_k y_2 - x_{k+1} y_1 - x_{k+2}|^2 \qquad (10)$$

Is minimum over all combinations of possible values of $x_k$, $x_{k+1}$ and $x_{k+2}$. Final the detector evaluates $c_{2,k}$ for each of the 16 combinations of values of $x_k$, $x_{k+1}$ and $x_{k+2}$, and takes as the detected values of $s_k$ of the data symbol of ,the value of $x_k$ corresponding to the smallest $c_{2,k}$.

## VI. SIMULATION RESULTS

The system shown in figure (1) has been built to transmit data along number of wired telephone and wireless channels as follows. Two different telephone circuits had been used to carry out the simulation, the attenuation and group delay characteristics are shown in Fig. 2a and Fig. 2b [7].

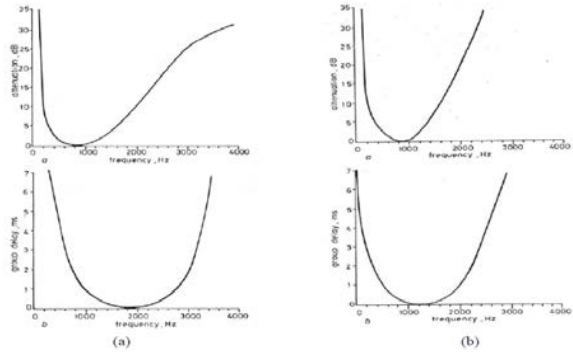

Fig .2. Attenuation and group delay characteristics of [7]: (a)Telephone circuit 1 (b)Telephone circuit 2

For sake of consistency with Fig .2a and Fig .2b, the attenuation and group delay characteristics of the equipment filters (transmission filter and receiver filter) are considered here. Table I shows the sampled impulse response of the (linear baseband channel) and (adaptive linear filter) shown in Fig .1 [7], also table II shows the sampled impulse response for telephone circuits 3, and 4.

TABLE I. SAMPLED IMPULSE RESPONSE FOR TELEPHONE CIRCUITS 1, AND 2 [7]

| Telephone circuit 1 | | Telephone circuit 2 | |
|---|---|---|---|
| *Real part* | *Imaginary part* | *Real part* | *Imaginary part* |
| 3.412 | 0.0667 | 0.2544 | 0.9941 |
| -0.13 | -0.0358 | -1.7394 | 0.2019 |
| 0.0263 | 0.0051 | 0.6795 | -0.8086 |

TABLE II. SAMPLED IMPULSE RESPONSE FOR TELEPHONE CIRCUITS 3, AND 4 [9]

| Telephone circuit 3 | | Telephone circuit 4 | |
|---|---|---|---|
| *Real part* | *Imaginary part* | *Real part* | *Imaginary part* |
| 0.4861 | 1.0988 | 0.5031 | 0.2008 |
| -0.598 | 0.0703 | -0.1447 | -0.0083 |
| 0.1702 | -0.1938 | 0.03 | -0.0097 |

## VII. SIMULATION TESTS

The BER of the various simulation programs has been measured by transmitting very large number of bits, and checking the received bits for errors, the BER is given by:

$$BER = \frac{\text{Number of Erroneous Bits}}{\text{Total Number of Bits}} \quad (11)$$

The BER should be calculated at different signal to noise ratios. The signal to noise ratio (SNR) in dB is given by equation (12),

$$SNR = 10\log_{10}\left(\frac{P_S}{N_N}\right) \quad (12)$$

Where PS represents the signal power and PN represents the noise power. The criteria for comparing the complexity of the simulated algorithms is based here on the number of computations performed per iteration, where the iteration is a sequence of processes done by the processor of the detector repeated every time a new symbol is received, generally [10], So, in this project, the algorithm with minimum amount of these operations per iteration among a set of algorithms under consideration will be considered as the one with the minimum complexity.

## VIII. SIMULATION OF THE IMPROVED NONLINEAR EQUALIZER WITH PERTURBATION

The existing algorithms can be simulated as follows, where the improved nonlinear equalizer with perturbation algorithm had been constructed to simulate different types of this algorithm with different values of computational complexity, increasing the number of the detection of the data symbol $s_k$ is delayed until the receipt of $r_{k+1}$. the detector now has determined $s'_{k+1}$. This increasing is engaged with a performance improvement as well as an increase in computational complexity, as shown earlier. The improved nonlinear equalizer has been tested with two types of channels and with the delay in the detection of n sampling interval by n=1, 2. Where the number of the delay in the detection of n sampling increased, this is led to increase the performance of the system, but calculation gives an indication about the complexity since it involves the largest number of multiplications for each received signal. This test had been carried out with two telephone channels as shown in Fig .2.

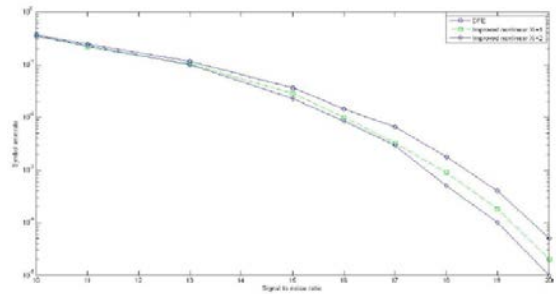

(a)

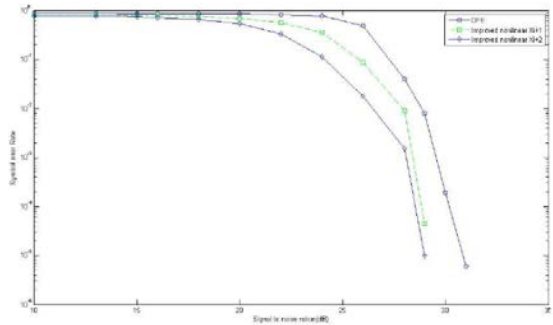

(b)

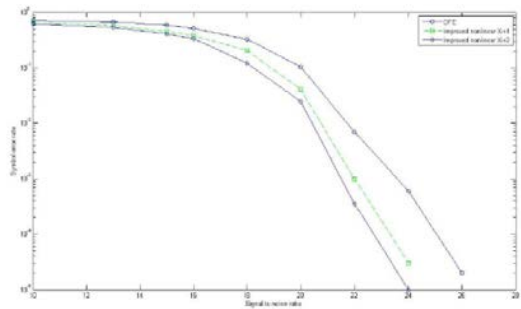

(c)

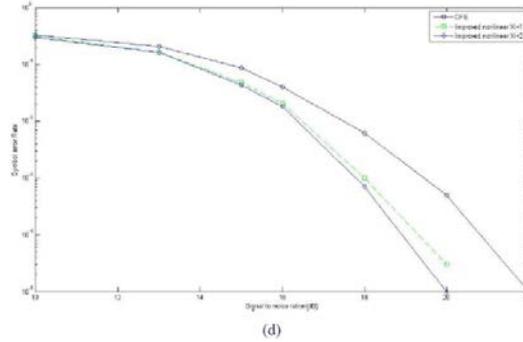

(d)

Fig .3. Comparison between various improved non linear equalizer with DFE under: (a) Telephone circuit 1. (b) Telephone circuit 2. (c) Telephone circuit 3. (d) Telephone circuit 4.

$x_{k+1}$, and $x_{k+1,k+2}$ as shown in Fig .3 represent the detected symbol with sample delay 1, and 2. The improved nonlinear equalizer of delay one, and two samples ($x_{k+1}$, $x_{k+1,k+2}$) has an advantage in tolerance to additive white Gaussian noise over the Decision feedback equalizer of about (0.5 and 1) dB, (1 and 2 dB), (1.5 and 2 dB), and (2 dB) respectively, for telephone circuits 1, 2, 3 and 4 respectively. This advantage can be seen that at error rate of around 1 in 102.

## IX. SIMULATING OF NEAR MAXIMUM LIKELIHOOD SEQUENCE ESTIMATION

The procedure [10] starts with 16 stored vectors {Qi-1}. Each of these is first expanded to give the corresponding vector Pi and hence the corresponding vector Qi with the smallest cost. To each of the two vectors {Qi} with the smallest and second smallest costs of the set of 16, are added three vectors {Qi} differing only in the last component and with the smallest costs. Then to each of the two vectors {Qi} with the third and fourth smallest costs of the original set of 16, is added the vector Qi differing only in the last component and with the smallest cost. No vectors are added to the four vectors {Qi} having the 5th to 8th smallest costs of the original set of 16 and the remaining eight of this set of vectors are discarded.

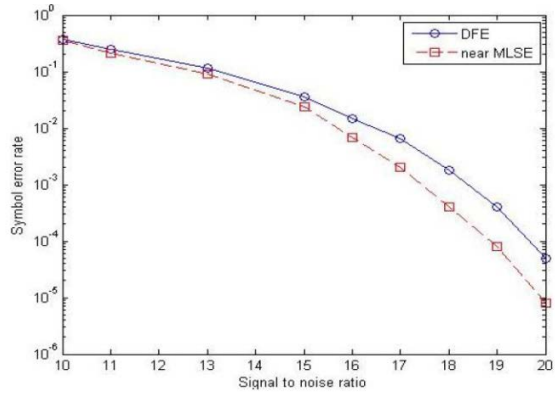

(a)

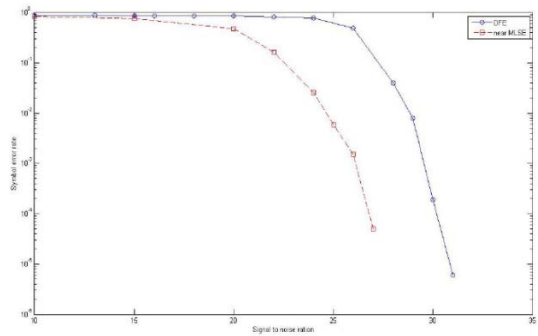

(b)

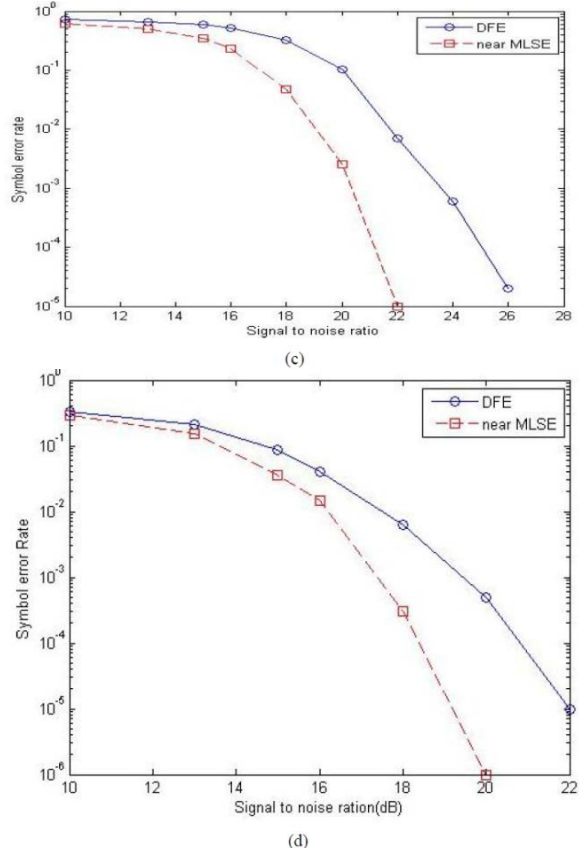

Fig .4. Performance of near MLSE when operating on (a) Telephone circuit 1 (b) Telephone circuit 2 (c) Telephone circuit 3 (d) Telephone circuit 4

The performance of this type of Near MLSE for telephone circuits 1, 2, 3, and 4 are shown in Fig .4a, 4b, 4c, and Fig .4d respectively. It can be seen that at error rate of around 1 in 103 , Near MLSE system has an advantage in tolerance to additive white Gaussian noise over the Decision feedback equalizer of about 2.5 and 4 dB, respectively, for telephone circuits 1 and 2 respectively. These results match the published results in [10].

X. SIMULATION RESULTS OF MOBILE CHANNEL

Simulation results for near maximum likelihood sequence estimation, decision feedback equalizer, and improved nonlinear equalizer with perturbation had been compared together under a multipath fading channel with Rayleigh distribution with two cases of mobile speed (3 km/h and 60 km/h), and a multipath spread of 0.25x10-6 sec, where minimum phase filter had been used with mobile channel.

The GSM model channel is shown in Fig .5, where the input signal $s_i(t)$ is passed through a tapped delay line which has a number of taps. Each tap represents a received signal path and the signals of all taps are summed together to produce the received signal. The delay blocks $T_1$ to $T_n$ represent the time delays between different paths while $Z_0$ to $Z_n$ ($Z=20 \log (4\pi d/\lambda)$) represent the paths attenuations, where d is the distance between the vehicle and BTS (500 m), $\lambda$ is the wavelength (0.333 m). Each tap (path) can be set to fade independently by a Rayleigh function $R_0$ to $R_n$ (-5.7 dB) (perform by a Rayleigh fading simulator), where these parameters can be set by the following equation.

$$r_i = \sum_{n=0}^{L} Z_n R_n s_{i-n} + w_i \qquad (13)$$

These functions ($R_0$ to $R_n$) simulate the Rayleigh fading phenomena results from the local reflections and diffractions near the mobile station.

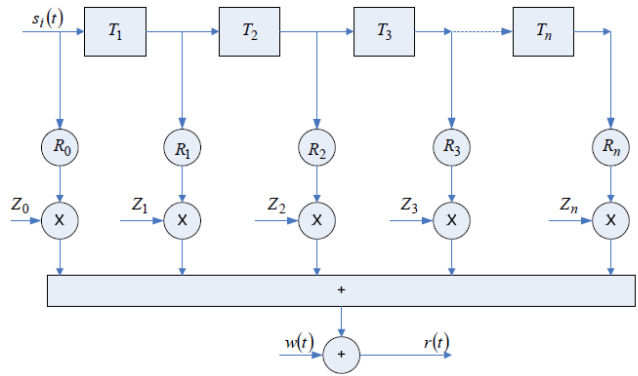

Fig .5. GSM channel model

Each path has different fade due to the different objects that cause the local reflections near the mobile station.

The comparison of the Near MLSE, decision feedback equalizer, and improved nonlinear equalizer had been carried out for mobile channel as shown in Figures .6, and Fig .7.

As shown in Fig .6, the improved nonlinear equalizer of one sample delay $x_{i+1}$ has advantage in tolerance to additive white Gaussian noise over DFE about 0.9 dB, but the near MLSE has advantage in tolerance to additive white Gaussian noise over DFE about 4.2 dB.

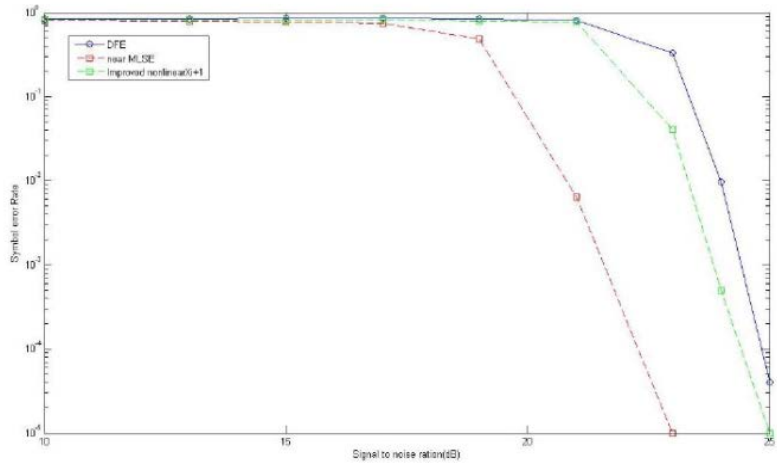

Fig .6. Performance comparison between near MLSE, DFE, and improved nonlinear equalizer of mobile station speed 3km/h.

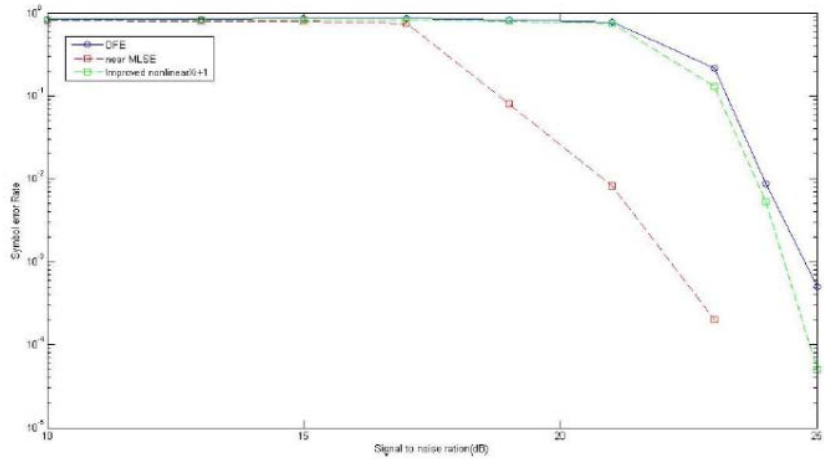

Fig .7. Performance comparison between near MLSE, DFE, and improved nonlinear equalizer of mobile station speed 60km/h.

For Fig .7, the improved nonlinear equalizer of one sample delay $x_{i+1}$ has advantage in tolerance to additive white Gaussian noise over DFE about 0.5 dB. The near MLSE has advantage in tolerance to additive white Gaussian noise over DFE about 2.7 dB.

## XI. THE RESULTS COMPARING

The comparing had been done between Near MLSE and Improved nonlinear The Results comparing equalizer under four telephone circuit channels. Where the Near MLSE algorithm used in this project is (16-8-16), in this arrangement, the algorithm calculates 24 costs each time a new symbol is received. The improved nonlinear equalizer used in this project of delay of one and two samples. The effective delay of one sampling interval calculated 16 different values of $c_{1,k}$ each time a new symbol is received. The effective delay of two sampling interval calculated 256 different values of $c_{2,k}$ each time a new symbol is received.

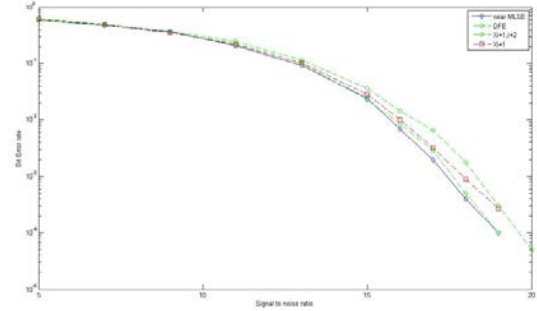

(a)

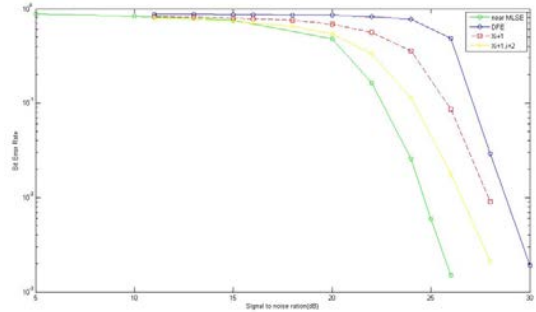

(b)

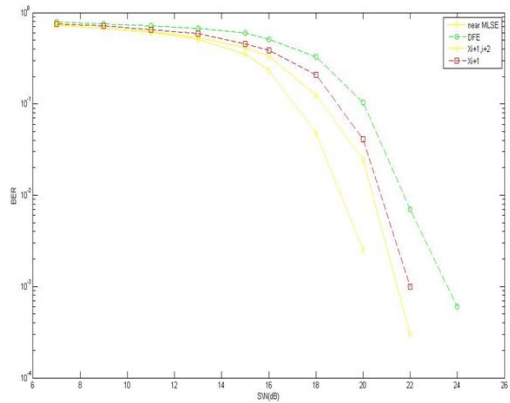

(c)

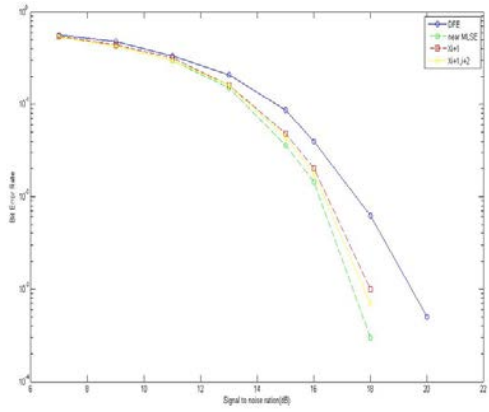

(d)

Fig .8. Comparison among various improved nonlinear equalizer, Near MLSE, and DFE under: (a) Telephone circuit 1. (b) Telephone circuit 2. (c) Telephone circuit 3. (d) Telephone circuit 4.

According to Fig .8a, 8b, 8c, and Fig .8d, the improved nonlinear has no advantage in the performance over the near MLSE, but has advantage in the performance over DFE where the complexity of improved nonlinear is higher than DFE .The complexity of Near MLSE is higher that the improved nonlinear equalizer. In focusing in the figures, it is clear that the performance of the improved nonlinear equalizer becomes better when the number of sampling delay is increasing, but the increasing of the number of sampling delay leads to more computations such that more complexity for the system, this is for telephone circuits' channels.

Fig .6, and Fig .7 show that the improved nonlinear equalizer of one sample delay has no advantage in the performance over the near MLSE, but has advantage in the performance over decision feedback equalizer, this advantage expresses a benefit for the improved nonlinear equalizer of one sample delay, where the complexity of improved nonlinear is higher than DFE. The complexity of Near MLSE is higher that the improved nonlinear equalizer. It is obvious from the Figures, that the performance of the improved nonlinear equalizer becomes better when the number of sampling delay is increasing, but the increasing of the number of sampling delay leads to more computations such that more complexity for the system, this is for mobile channel.

Table III shows a comparison between various reduced complexity algorithms.

TABLE III. A COMPARISON BETWEEN VARIOUS REDUCED COMPLEXITY ALGORITHMS

| Equalization Method | Number of cost evaluated per data symbol | Number of separate operations (multiplication add, and sub) | Number of separate operations involved in the search | Number of total operations required per symbol |
|---|---|---|---|---|
| *Near Maximum Likelihood Detection* | 32 | 640 | 304 | 944 |
| *Improved nonlinear Equalizer with Perturbation of one sample delay* | 4 | 39 | 32 | 71 |
| *Improved nonlinear Equalizer with Perturbation of two sample delay* | 8 | 59 | 48 | 107 |
| *Decision Feedback Equalizer* | 1 | 19 | 16 | 35 |

## XII. CONCLUSIONS

The following conclusions could be made:

- Under telephone circuit channels 1, 2, 3, and 4, it is clear that the Near MLSE detector has an advantage in performance over the improved nonlinear equalizer. for the telephone circuit channel 1 the Near MLSE detector has advantage in performance rather more than the improved nonlinear equalizer ( $x_{k+1}$, $x_{k+1,k+2}$ ), for the channel 2 the Near MLSE detector has advantage in performance over the improved nonlinear equalizer( $x_{k+1}$, $x_{k+1,k+2}$ )by 2 dB, and 3 dB respectively. For channel 3 the Near MLSE has advantage in performance over the improved nonlinear equalizer ( $x_{k+1}$, $x_{k+1,k+2}$ ) by 1.5 dB, and 2 dB respectively. For channel 4 the Near MLSE has advantage in performance over the improved nonlinear equalizer ( $x_{k+1}$, $x_{k+1,k+2}$ ) by 2 dB, and 3 dB respectively. Among the above difference in performance between the Near MLSE and the improved nonlinear equalizer (all the fourth telephone circuit channel are minimum phase),the Near MLSE detector has more complexity more than the improved nonlinear equalizer by 20 % .

- Under minimum phase multipath fading channel with mobile speed of both 3Km/h, and 60Km/h; Near MLSE has more advantage in performance than the improved nonlinear equalizer with perturbation of one sample delay, also more computational complexity for the Near MLSE than the improved nonlinear equalizer with perturbation.

- The performance of the improved nonlinear equalizer with Perturbation increases when the number of sampling delay increases, it is obvious that the improved nonlinear equalizer of 1 sampling delay ( $x_{k+1}$ ) has less performance and less computational complexity than the improved nonlinear equalizer of 2 sampling delay.

ACKNOWLEDGMENT

The authors would like to thank Islamic Development Bank (IDB) and university of Ottawa, School of Information Technology and Engineering, Centre for Research in Photonics, Dr. Trevor and Dr. Sawsan for their support to complete this work.